\documentclass[aps,pre,
 onecolumn,
twocolumn,
10pt,
 notitlepage,
 showpacs,floatfix,nofootinbib,
 superscriptaddress
 ]{revtex4-1}

 \usepackage{subfigure}
 \usepackage{amssymb}
 \usepackage{amsfonts}
 \usepackage{amsmath}
 \usepackage{amsthm}
 \usepackage{epsfig}
 \usepackage{float}
 \usepackage{multirow}
 \usepackage[usenames,dvipsnames]{color}
 \usepackage[latin1]{inputenc}
 \usepackage{xr}
 \usepackage{enumitem}
 \usepackage{graphics,graphicx,xcolor,subfigure}
 \usepackage[hidelinks,unicode=true]{hyperref}
 \hypersetup{colorlinks=true,% Don't draw a box around links, instead color every piece of text, which is a link
 	linkcolor=blue,
 	urlcolor=blue,
 	citecolor=blue,
 	pdfhighlight=/N% When clicking a link and holding the mouse button, don't use any graphical effects - i.e. the alternative would be to behave like a "button" which is pressed within the PDF
 }%\usepackage{hyperref}
 
 \usepackage{comment}
 \usepackage[normalem]{ulem}

\newcommand{\av}[1]{\langle {#1} \rangle}

\newcommand{\revis}[1]{{\color{black} {#1}}}
\newcommand{\add}[1]{{\color{black} #1}}

\newcommand{\orcid}[1]{\hspace{0.2em}\href{https://orcid.org/#1}{\includegraphics[keepaspectratio,width=0.7em]{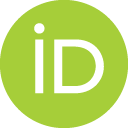}}}

\begin{document}

\title{Outbreak diversity in epidemic waves propagating through distinct geographical scales}

\author{Guilherme S. Costa\orcid{0000-0002-5019-0098}}
\affiliation{Departamento de F\'{\i}sica, Universidade Federal de Vi\c{c}osa, 36570-900 Vi\c{c}osa, Minas Gerais, Brazil}

\author{Wesley Cota\orcid{0000-0002-8582-1531}}
\affiliation{Departamento de F\'{\i}sica, Universidade Federal de Vi\c{c}osa, 36570-900 Vi\c{c}osa, Minas Gerais, Brazil}

\author{Silvio C. Ferreira\orcid{0000-0001-7159-2769}}
%\email{silviojr@ufv.br}
%\thanks{Corresponding author.}
\affiliation{Departamento de F\'{\i}sica, Universidade Federal de Vi\c{c}osa, 36570-900 Vi\c{c}osa, Minas Gerais, Brazil}
\affiliation{National Institute of Science and Technology for Complex Systems, 22290-180, Rio de Janeiro, Brazil}

\begin{abstract}
A central feature of an emerging infectious disease in a pandemic scenario is
the spread through  geographical scales and the impacts on different locations
according to the adopted mitigation protocols. We investigated a stochastic
epidemic model with the metapopulation approach in which patches represent
municipalities. Contagion follows a stochastic compartmental model for
municipalities; the latter, in turn, interact with each other through recurrent
mobility. As a case of study, we consider the epidemic of COVID-19 in Brazil
performing data-driven simulations. Properties of the simulated epidemic curves
have very broad distributions across different geographical locations and
scales, from states, passing through intermediate and immediate regions down to
municipality levels. Correlations between delay of the epidemic outbreak and
distance from the respective capital cities were predicted to be strong in
several states and weak in others, signaling  influences of multiple epidemic
foci  propagating towards the inland cities. Responses of different regions to a
same mitigation protocol can vary enormously implying that the policies of
combating the epidemics  must be engineered according to the region' specificity
but integrated with the overall situation. \revis{Real series of reported cases
	confirm the qualitative scenarios predicted in simulations.} Even though we
restricted our study to Brazil, the prospects and model can be extended to other
geographical organizations with heterogeneous demographic distributions.
\end{abstract}
 %\pacs{}
	
\maketitle

\section{Introduction}

Theoretical and computational toolboxes  used and developed by  physicists over
decades and, in particular, those of equilibrium and nonequilibrium statistical
physics have  found   an inexhaustible source of applications in complexity~\cite{Bak1988,Barabasi2005,Goldenfeld1999,Nowak1992} attaining an apex
with the raising of network science at late 1990s~\cite{barabasi2016network}.
Epidemic models, which are intimately related with nonequilibrium phase
transitions~\cite{Henkel2008}, running on the top of  networks soon became one of the
most imminent fields of interdisciplinary
physics~\cite{Pastor-Satorras2001,Newman2002,PastorSatorras2015} with many
applications on real epidemic scenarios lead by
physicists~\cite{Colizza2006,Balcan2009,Wang2009}. The pandemic of coronavirus
disease (COVID-19) has been marked by a wide discussion heavily grounded on
data-driven epidemic modeling, again with outstanding contributions lead by
physicists~\cite{Chinazzi2020,Kraemer2020,Gilbert2020,Arenas2020b}.

Since COVID-19 outbreak was reported in Wuhan, Hubei province, China, at the end
of December 2019, the scientific community has turned its attention to
understanding and  combating  this novel threat~\cite{Li2020a, Wu2020,
	Chinazzi2020, Li2020, Kraemer2020}. A characteristic of this infectious disease,
caused by the pathogen coronavirus SARS-CoV-2, is the high transmission rate of
individuals \add{ without significant symptoms}~\cite{Li2020,Read2020} who
traveled unrestrictedly and spread COVID-19 through all continents except
Antarctica. Efforts to track the transmission from Wuhan to the rest of mainland
China and world relied on mathematical modeling~\cite{Chinazzi2020, Kraemer2020,
	Pullano2020, Wu2020, Aleta2020a}, which were soon extended to other
countries~\cite{Danon2020,Arenas,Aleta2020c}, using the metapopulation
approach~\cite{Colizza2007a,Danon2009,Gomez-Gardenes2018}. In this modeling, the
population is grouped in patches representing geographic regions and the
epidemic contagion obeys standard compartmental models~\cite{diekmann2000}
within the patches, while mobility among them promotes the epidemic to spread
through the whole population. Metapopulation approaches, which were initially
conceived in ecology contexts~\cite{Hanski1989}, have been used for many
epidemic processes~\cite{Colizza2007,Colizza2007a, Watts2005,Lloyd2004,
	Danon2009, Gomez-Gardenes2018, Meloni2011} and allow  a  geographical resolution
absent in standard compartmental models~\cite{diekmann2000}. One remarkable
feature of the metapopulation approach is its capability to integrate the
epidemic modeling in different regions into a single framework, in which
correlation between epidemic waves in different regions can be investigated.

We are interested on the   geographical variability and impacts of an emerging
and highly contagious infections disease such as the COVID-19 in continental
scales and we chose the recent pandemics in  Brazil as a case of study. Brazil
is a country of continental territory with an area of $8.5\times10^6$~km$^2$
demographically, economically, and developmentally very heterogeneous. It is a
federation divided into 26 states and one Federal District, its capital city.
States are divided into municipalities; each state has its own capital city. The total
number of municipalities is 5570 and, according to estimates of 2019,  the 
country population is approximately 210 million~\cite{ibge2019estimativas}. Striking
characteristics of the Brazilian demography are the broad distributions of the rural
and urban population fractions and urban population densities of municipalities, which are important
traits in epidemic modeling~\cite{Hu2013,Rocklov2020}.
Therefore, investigation of a pandemic  and its consequences have necessarily to
integrate the particularities of each region, which can be reckoned with the
metapopulation approach.

In this work, we investigate a metapopulation model and applied it to the
COVID-19 epidemic spread in the Brazilian municipalities. \add{At the time of
	submission, early June 2020,} the benchmark of 800~000
cases and 40~000 deaths was surpassed in Brazil, which became in this date the
country with highest epidemic incidence and possibly a new worldwide epicenter
of COVID-19~\cite{covidBR}.  \add{At time of resubmission, mid October 2020, the
	number of cases had surpassed 5.5 millions with more than 150~000 deaths.} The model
is supplied by demographic and mobility data mined from public sources, which
are used in the stochastic simulations. We  quantify how diversified the
epidemic can be across the country rather than try to accurately foresee
outbreaks in specific locations. We  determine the level of dispersion in the
curves of the epidemic prevalence in different places within a hypothetical
scenario of uniform mitigation measures. We take the situation of 31 March 2020
as the initial condition, with foci in 410 municipalities, allowing to track the
epidemics in all municipalities in a single integrated simulation. We
predicted that the prevalence curves can vary enormously over
geographical scales ranging from states down to municipalities. We also
predicted that some states are characterized by outbreaks
starting mainly in their capital cities, followed by epidemic waves propagating
toward the countryside, while in other states  there are multiple epidemic foci.
Our results explain why policies of combating a countrywide pandemic should be
simultaneously decentralized, in the sense that each locality has to consider
its specificities, and integrated, since the fate of one region depends on the
attitudes adopted by others. \add{These qualitative behaviors have been
	confirmed \textit{a posteriori.} }

The remaining of the paper is organized as follows. The model is described in
Sec.~\ref{sec:model} while its computer implementation and parameters are in
Appendixes~\ref{app:model} and \ref{app:par}, respectively.  Model calibration
using nowcasting is presented in Sec.~\ref{sec:now} while the detailed results
are in Sec.~\ref{sec:results}. Some discussions and remarks are presented in
Sec.~\ref{sec:discu}. \add{In Sec.~\ref{sec:after}, we present a brief afterwords
	commenting the achievements of the model in forecasting the qualitative countrywide
	behavior of the  pandemics in Brazil five months after the predictions were
	made~\cite{Costa2020}.}

\section{The model}
\label{sec:model}

\subsection{Metapopulation}
\begin{figure}[t]
	\centering
	\includegraphics[width=0.99\linewidth]{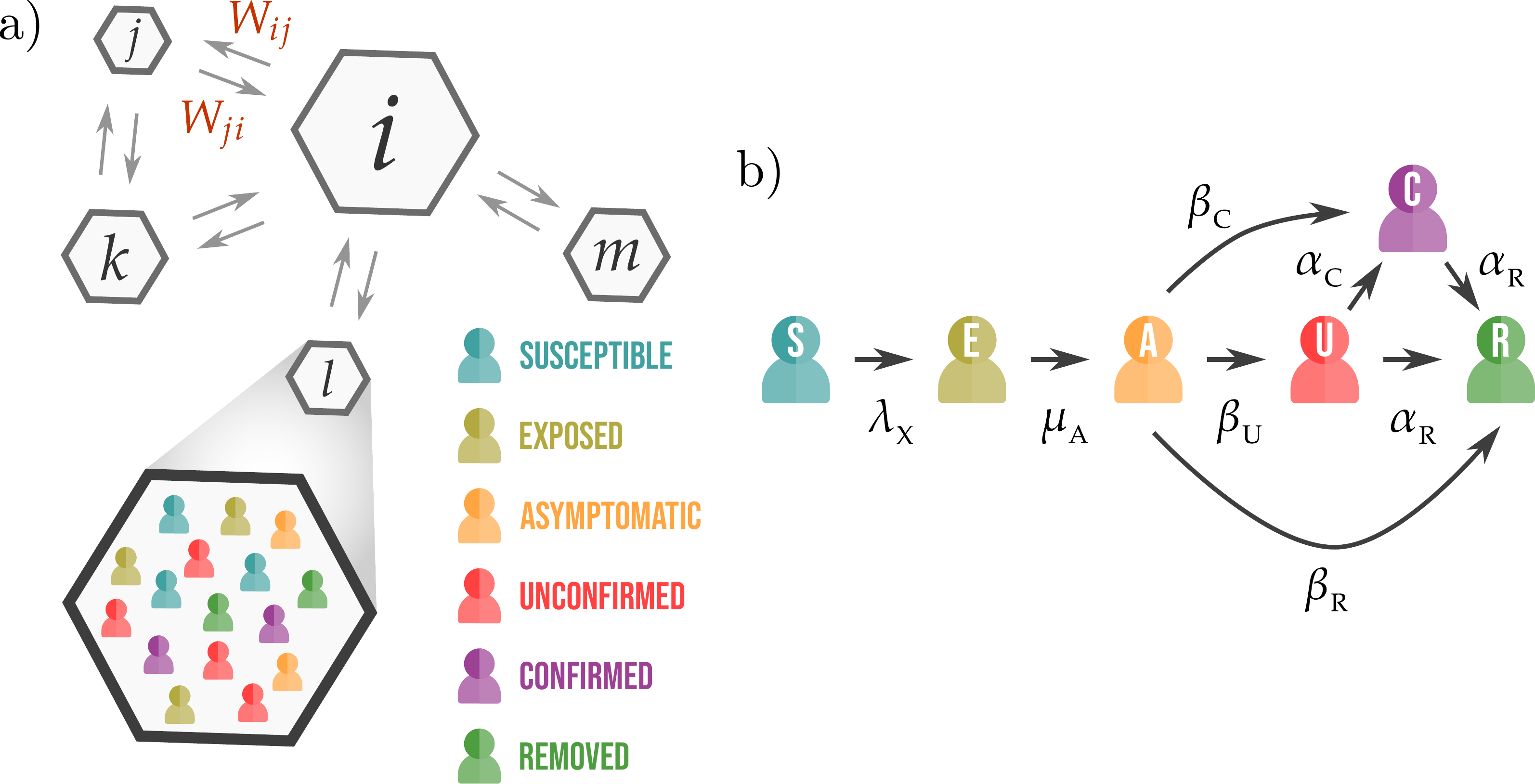}
	\caption{Schematic representation of  a) metapopulation structure  and b)
		epidemic transitions with their respective rates. Patches $i$ and $j$ of
		different population sizes (represented by the areas) can interchange
		individuals with rates $W_{ij}$ and  $W_{ji}$. The epidemic events occur within
		patches under the hypothesis of well-mixed populations as illustrated in the
		zoom of patch $l$. Six compartments are used: susceptible (S), exposed (E),
		asymptomatic \add{or presymptomatic} (A), unconfirmed (U), confirmed (C), and removed (R). The
		infection rate per contact $\lambda_\text{X}$ can be of three forms, depending
		on the infectious contacts: S and A ($\lambda_\text{A}$), S and U
		($\lambda_\text{U}$), and S and C ($\lambda_\text{C}$).}
	\label{fig:model_scheme}
\end{figure}

We consider a stochastic metapopulation model for the countrywide  spread of the
epidemic, in which a key ingredient is recurrent
mobility~\cite{Gomez-Gardenes2018} whereby individuals travel to other patches
but return to their residence  recurrently. The population is divided into
patches which represent the places of residence  as schematically shown in
Fig.~\ref{fig:model_scheme}a). Similar models have been used for modeling of
COVID-19 in Spain~\cite{Arenas,Arenas2020b,Aleta2020c}. In particular, our model
is inspired on a recent work~\cite{Arenas} to investigate the COVID-19. We  use
a different, continuous-time approach where  geographical, demographical, and
testing capacity specificities are included. As in general metapopulation
models, a patch can represent a municipality, a neighborhood or any other
demographic distribution of interest. Each patch is labeled by
$i=1,2,\ldots,\Omega$, where $\Omega$ is the number of patches, and has a
resident population fixed in time given by $N_i$. The inter-patch
 mobility rate is given by $W_{ij}$ defined as the number of
individuals that travel from $i$ to $j$ and return to $i$ after a typical time
$T_\text{r}$. A patch $j$ for which $W_{ij}>0$ is a \textit{neighbor} of $i$.
Different forms of mobility, with their corresponding characteristic times, can
be implemented. For the sake of simplicity,  recurrent flux  of individuals  per
day ($T_\text{r}=1$~d) is used in the present work. Mobility within patches is
not explicitly considered. Instead, we assume the well-mixed population
hypothesis~\cite{diekmann2000} where an individual has equal chance to be in
contact with any other in the same patch.

\subsection{Demography and contacts}
The average number of contacts per  individual of patch $i$ per unit of time is
$k_i$. Variations in the social distancing can be implemented by altering $k_i$.
Contacts are related to local mobility and are strongly correlated with  the
population density $\xi_i$. Monotonic increase has been proposed~\cite{Hu2013}.
We follow Ref.~\cite{Arenas} and use a contact number modulated by $k_i\propto
f(\xi_i)$ with $f(\xi)=2-\exp(-a\xi)$. To take into account the sparse
demographic distribution of Brazil, we used the urban population density of each
municipality to compute $\xi_i$. Interactions between urban and rural populations
are also explicitly considered through a symmetric $2\times 2$ contact matrix
$\mathcal{C}_\text{uu}$, $\mathcal{C}_\text{rr}$, and
$\mathcal{C}_\text{ur}=\mathcal{C}_\text{ru}$, $0\le \mathcal{C}_\text{xy}\le 1$,
determining the  contacts between urban (u) and rural (r) populations whose
fractions  are represented by $\omega_i$ and $1-\omega_i$, respectively. The
number of contacts in patch $i$  is also modulated by
\begin{equation}
k_i\propto g_i=
\mathcal{C}_\text{uu}\omega_i^2 + 2\omega_i(1-\omega_i)\mathcal{C}_\text{ur}+
(1-\omega_i)^2 \mathcal{C}_\text{rr},
\end{equation}
where the terms  represent urban-urban, urban-rural, and rural-rural interactions
from left to right. The contacts are parameterized  by their average  number per
individual $\av{k}$, such that $k_i=zg_if(\xi_i)\av{k}$, where $z$ is a
normalization factor, required by the relation $\sum_i k_i N_i =
N_\text{tot}\av{k}$ and given by 
\begin{equation}
z=\frac{N_\text{tot}}{\sum_{i=1}^{\Omega} f(\xi_i)g_i N_i}, 
\end{equation}
where $N_\text{tot}= \sum_i N_i$ is the total population.

\subsection{Epidemic model}

The epidemic dynamics within the patches considers three noncontagious compartments
which are: susceptible (S), that can be infected; removed (R), that represents
immunized or deceased individuals; and exposed (E) that carries the virus but
does not transmit it. Compartments of contagious individuals are: asymptomatic
(A), \add{that have not presented
	symptoms}; unconfirmed (U), which are
symptomatic but not yet identified by tests; and confirmed (C), that have tested
positive for COVID-19. 
\add{The compartment A is actually a simplification of the real
	situation. At least two relevant types of asymptomatic persons do exist:
	presymptomatic, that will develop symptoms at the end of the incubation period,
	and the truly asymptomatic, that becomes recovered without any significant
	symptoms. There exist large uncertainties with respect to the fraction of
	individuals who are truly asymptomatic~\cite{Byambasuren2020,Casey2020}.
	According to the parameters used in this work (see Appendix~\ref{app:par}), we
    have approximated 30\%  of truly asymptomatic in compartment A which lies within
	the interval reported elsewhere~\cite{Byambasuren2020}. Due
	to the uncertainty of the available data, we also assume a same contagion rate for  both types of
	asymptomatic individuals which are hereafter treated as a single compartment.}
The epidemic transitions obey a SEAUCR  model,  in which susceptible individuals
become exposed  by contact with contagious persons (A, U, or C) with rates 
$\lambda_\text{A}$, $\lambda_\text{U}$, and $\lambda_\text{C}$, respectively.
Exposed individuals spontaneously become asymptomatic with rate $\mu_\text{A}$,
while the latter transitions to unconfirmed \add{(presymptomatic)}, confirmed, or
removed \add{(truly asymptomatic)} states  happen with rates $\beta_\text{C}$,
$\beta_\text{U}$, and $\beta_\text{R}$, respectively. Finally, unconfirmed or
confirmed individuals are removed from the epidemic process (via recovery or
death) with rate $\alpha_\text{R}$.  The transition U$\rightarrow$C has rate
$\alpha_{C}$ that may depend on the number of unconfirmed cases $n_i^\text{U}$. For
a small $n_i^\text{U}$ the confirmation rate is constant
$\alpha_\text{C}^{(0)}$, assuming that any municipality has the minimal
resources for testing. The number of confirmed cases is limited by the finite 
testing capacity per inhabitant, $\zeta$. We assume a simple monotonic function
\begin{equation}
\alpha_\text{C} = \frac{\alpha_\text{C}^{(0)}}{1+\frac{\alpha_\text{C}^{(0)} 
		n_i^\text{U}}{\zeta N_i}}.
\end{equation}
The transition E$\rightarrow$U  was not included since most studies agree that
the asymptomatic or presymptomatic condition is a central characteristic of
COVID-19~\cite{Li2020,He2020,Casey2020,Byambasuren2020}. The contact of confirmed individuals is
reduced to $bk_i$ with $b<1$. Figure~\ref{fig:model_scheme}b) shows a schematic
representation of the epidemic events.

Individuals residing in $i$ can move to neighbors $j$ following a stochastic
process with rates $R_i/T_\text{r}$ where $R_i$ is the population fraction of
$i$ that moves recurrently; the destination $j$ is chosen proportionally to
$W_{ij}$. Symptomatic (U or C) individuals do not leave their residence patches
but if the transitions to U or C happen out of his/her residence  they return
with rate $1/T_\text{r}$, as do the individuals in other epidemic states.
Infection, healing  and other transitions follow Poisson processes with the
corresponding rates. Details of the computer implementation, using an optimized
Gillespie algorithm~\cite{Cota}, are given in  the  Appendix~\ref{app:model}.

\subsection{Initial condition}

\begin{figure}[!t]
	\centering
	\includegraphics[width=0.95\linewidth]{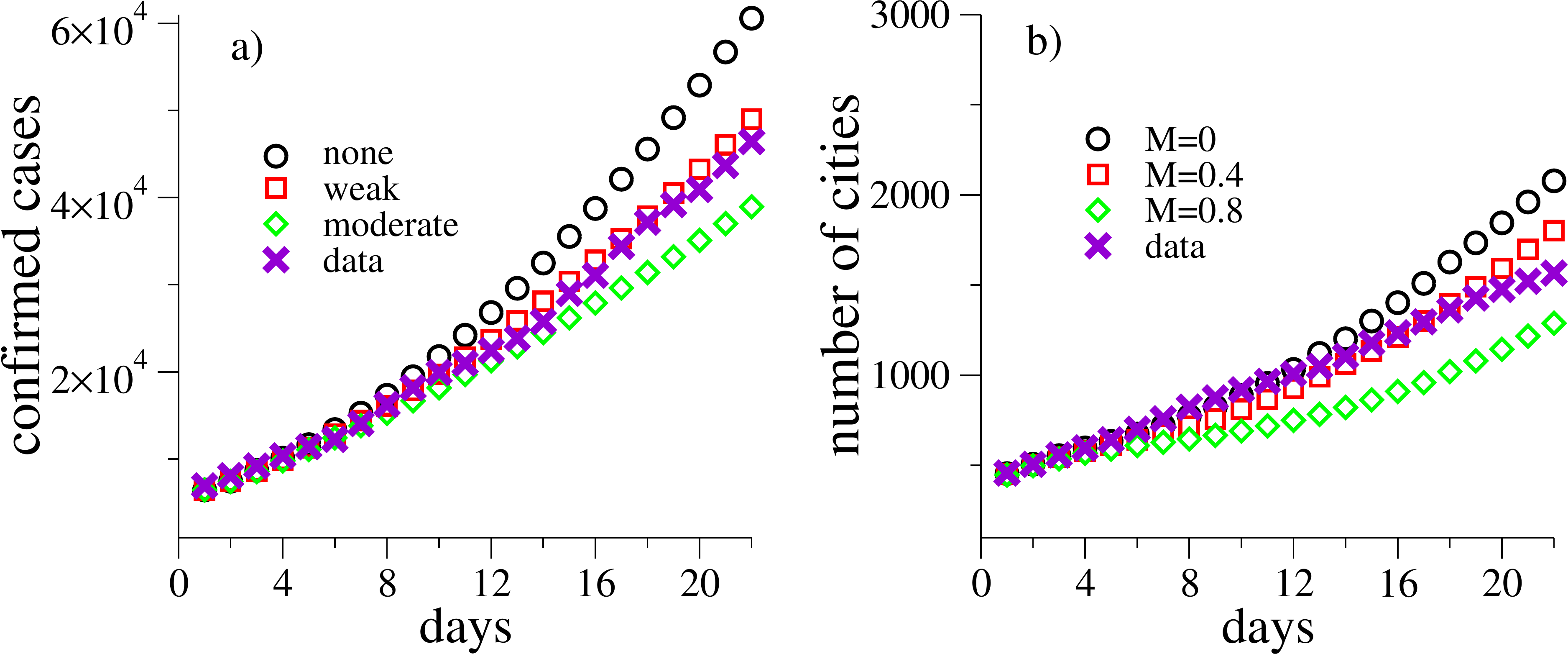}
	\caption{Comparison  of simulations with reported data for COVID-19 in
		Brazil~\cite{covidBR} for the first three weeks with day 0 corresponding to 31
		March 2020. a) Number of  confirmed cases are presented for different mitigation
		scenarios: none with $(M,K)=(0,0)$, weak  with $(M,K)=(0.4,0.3)$
		and moderate with $(M,K)=(0.8,0.5)$. b) Evolution of the number of municipalities
		with confirmed  cases for a reduction of contact of $K=0.3$ and different levels
		of long distance mobility given by $M=0,~0.4$, and $0.8$.}
	\label{fig:brasil_compare}
\end{figure}

The initial condition is a major hurdle for the accurate forecasting of a
pandemic  due to high underscore and delays in reporting. {First studies of}
the COVID-19 Wuhan outbreak estimated that undocumented
infections were the  source of at least 79\% of documented cases~\cite{Li2020}.
This number can be larger depending on the testing
policies~\cite{Bendavid2020,RSamostragem}. \add{The investigated scenario was
	the beginning of the pandemics in Brazil when testing policies and data
	availability at municipality level were still very premature.} We assume a simple
linear correlation between the increase in the number of  confirmed cases of
COVID-19 reported by the official authorities and infected cases (E, A, and U)
in each municipality. We used 31 March  2020 as the initial condition. Let
$\tilde{n}_i^\text{conf}(t)$ and $\tilde{n}_i^\text{rem}(t)$ be the number of
confirmed and removed (recovered plus deceased) cases of COVID-19 in each
municipality $i$~\cite{covidBR} of Brazil at day $t$. Let $t_{-1}$ correspond to
24 March, $t_0$ to 31 March, and $t_1$ to 4 April 2020. These data are provided
in supplementary material (SM)~\cite{supp}. The initial condition at $t_0$  is
\begin{eqnarray}
\tilde{n}_i^\text{C}(t_0) &  = & \tilde{n}^\text{conf}_{i}(t_0) -\tilde{n}^\text{rem}_{i}(t_0)  \nonumber \\
\tilde{n}_i^\text{U}(t_0) & = &\tilde{n}^\text{conf}_{i}(t_1)-\tilde{n}^\text{conf}_{i}(t_0) \nonumber\\
\tilde{n}_i^\text{A}(t_0) & = &2\left[\tilde{n}^\text{conf}_{i}(t_1)- \tilde{n}^\text{conf}_{i}(t_0)\right]\nonumber\\
\tilde{n}_i^\text{E}(t_0) & = &8\left[\tilde{n}^\text{conf}_{i}(t_1)- \tilde{n}^\text{conf}_{i}(t_0)\right] \nonumber\\
\tilde{n}_i^\text{R}(t_0) & = & 12 \tilde{n}^\text{conf}_{i}(t_{-1}),
\end{eqnarray}
where $\tilde{n}^\text{X}_{i}$  is the number of individuals in patch $i$ and
state X. The last equation means that essentially all  individuals who were
infected one week before are currently removed. Since we are far from herd
immunity at $t_0$, this choice is irrelevant. This approach, based on the
hypothesis that the increase of reported cases is  correlated with local
transmission, is a way to reduce the effects of underreporting. The prefactors
were chosen together with the parameter $\alpha_\text{C}^{(0)}$ to fit both  the
initial increase (first two weeks of March) of  the total number cases and of
municipalities with confirmed cases under a weak mitigation scenario.
%The confirmation rate $\alpha_\text{C}^{(0)}$, the ratio between times in
%exposed and asymptomatic compartments, and  the initial conditions, which are
%the unknown epidemiological  parameters of the model, were calibrated to reproduce
%the total number of confirmed cases  and the number  of municipalities with
%confirmed cases  for Brazil in the {first three weeks} of April under a weak
%mitigation scenario. 
The last one is consistent with the estimates of the \textit{Google Community
	Mobility Reports} for Brazil within this time window~\cite{aktay2020google}; see
Figs.~\ref{fig:brasil_compare} and \ref{fig:Ncity_UF2}.

\subsection{Mitigation strategies}
Different mitigation strategies are studied by uniformly reducing
the \add{inter-patch} mobility and contact by  factors $(M,K)$,
i.e., $W_{ij}\rightarrow (1-M)W_{ij}$, for $j\neq i$, and $k_i\rightarrow (1-K)
k_i$. We investigate three different scenarios: none, weak, and moderate
mitigations represented by $(M,K)=(0,0)$, $(0.4,0.3)$, and $(0.8,0.5)$,
respectively. We do not tackle the regime of suppression or strong mitigation
that could drop the epidemic drastically. Averages were evaluated over 10
independent simulations.

\subsection{Remarks on parameter choices}
The extensive set of parameters is aimed to reproduce the essence of the
epidemic spreading at a countrywide level but not to track accurately the
epidemic outbreak in every state or municipality. We specially assume uniform
and constant control parameters for reduction of
\add{inter-municipality} movement and contacts as well as the
testing rates. This is surely not the real case.  Each federative state or
municipality has its own mitigation  and testing policies that are very
heterogeneous. So, assuming uniformity and fitting the overall data at a country
level, it is possible to infer in which regions the epidemic is evolving faster
or slower than the average. Finally, our initial condition is a guess driven by
data but still  limited. The level of underreporting is known to be high, and
broad ranges of values have been
reported~\cite{Li2020,Bendavid2020,RSamostragem}. The description of the
complete set of epidemiological parameters, the initial conditions, mobility and
demographic  data are given in Appendix~\ref{app:par}.

\section{Nowcasting and model calibration}

\label{sec:now}

\begin{figure*}[t]
	\centering
	\includegraphics[width=0.95\linewidth]{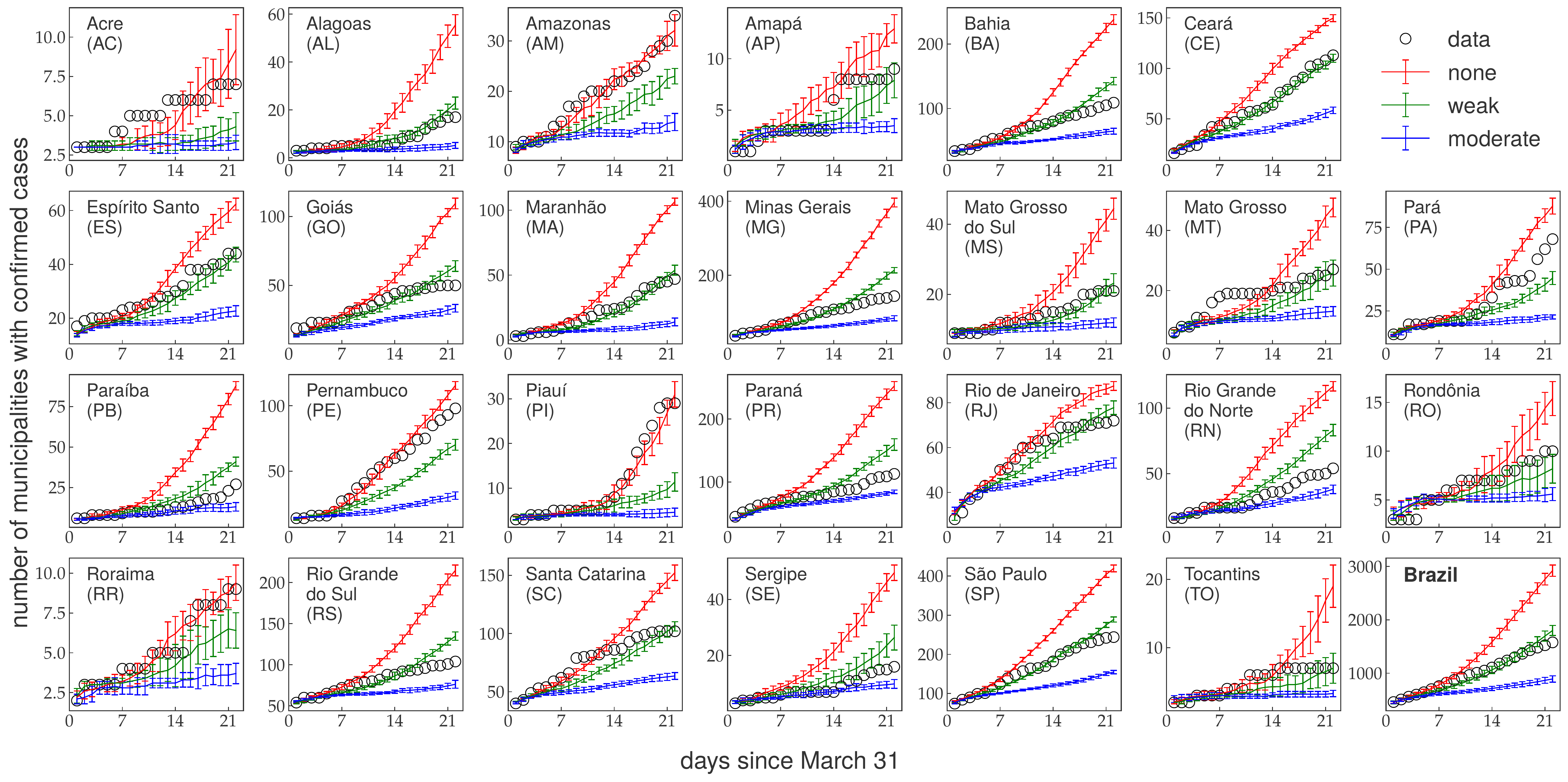}
	\caption{Simulations and observations of the number of  municipalities with
		confirmed cases of COVID-19 for the 26 federative states of Brazil in the first
		three weeks of simulations. Data for the whole country are shown in  the bottom
		right corner. Simulations were performed using none, weak, and moderate
		mitigations with parameters $(M,K)=(0,0),~(0.4,0.3)$, and $(0.8,0.5)$,
		respectively.}
	\label{fig:Ncity_UF2}
\end{figure*}

Irrespective of the heterogeneous testing approaches among
states~\cite{covidBR}, we assume that essentially all municipalities
{had} capacity to detect COVID-19 in the first hospitalized
patients and, therefore, the number of municipalities with positive cases is
expected to be an observable much less prone to the effects of underreporting
than the number of confirmed cases \add{at the beginning of the pandemics in
	Brazil}. Figure~\ref{fig:Ncity_UF2} compares the evolution of municipalities
with confirmed cases  between 1 and 22 April 2020 obtained in simulations with
different mitigation measures and real data  for each Brazilian state and also
for the whole country. Weak mitigation provides the best agreement with the
reported data whereas deviations start to appear after three weeks. The weak
mitigation parameters $(M,K)=(0.4,0.3)$, which fit better the overall data for
Brazil, also perform well for most of the  states in this time window and
especially those with higher incidence of COVID-19 {in the period}, which
lead the averages over the country. However, some states  (AC, AM, PA, PE, PI,
and RR - abbreviations for Brazilian  federative states are given in
Fig.~\ref{fig:Ncity_UF2}) {were} more consistent with the
simulations without mitigation while  others (PB, PR, and RN) were nearer to a
moderate mitigation. No tuning with respect to specific places was used and,
consequently, it does not reproduce accurately the number of reported cases for
all federative states due to the diversified testing and social distancing
policies across different states {and municipalities}, in contrast with the
{uniformity} hypothesis of the present study.

\section{Geographical analysis of epidemic prevalence}
\label{sec:results}

The role of \add{inter-municipality} traffic  is
investigated reducing the contacts by 30\% and varying the  mobility. The total
number of cases changes very little since the leading contribution comes from 
local transmission in municipalities with larger incidence  while the number of
municipalities with confirmed cases, shown in Fig.~\ref{fig:brasil_compare}b),
depends significantly on the \add{inter-municipality}
traffic restrictions. These results confirm the  epidemiological concept that
reducing long-distance travelling contributes to delaying the outbreak onset in
different places, but  alters  little the countrywide epidemic
prevalence~\cite{Colizza2007b}, as  discussed  in the context of
COVID-19 spreading in China~\cite{Chinazzi2020,Du2020} and
Spain~\cite{Aleta2020c}, for example.

\begin{figure*}[t]
	\centering
	\includegraphics[width=0.97\linewidth]{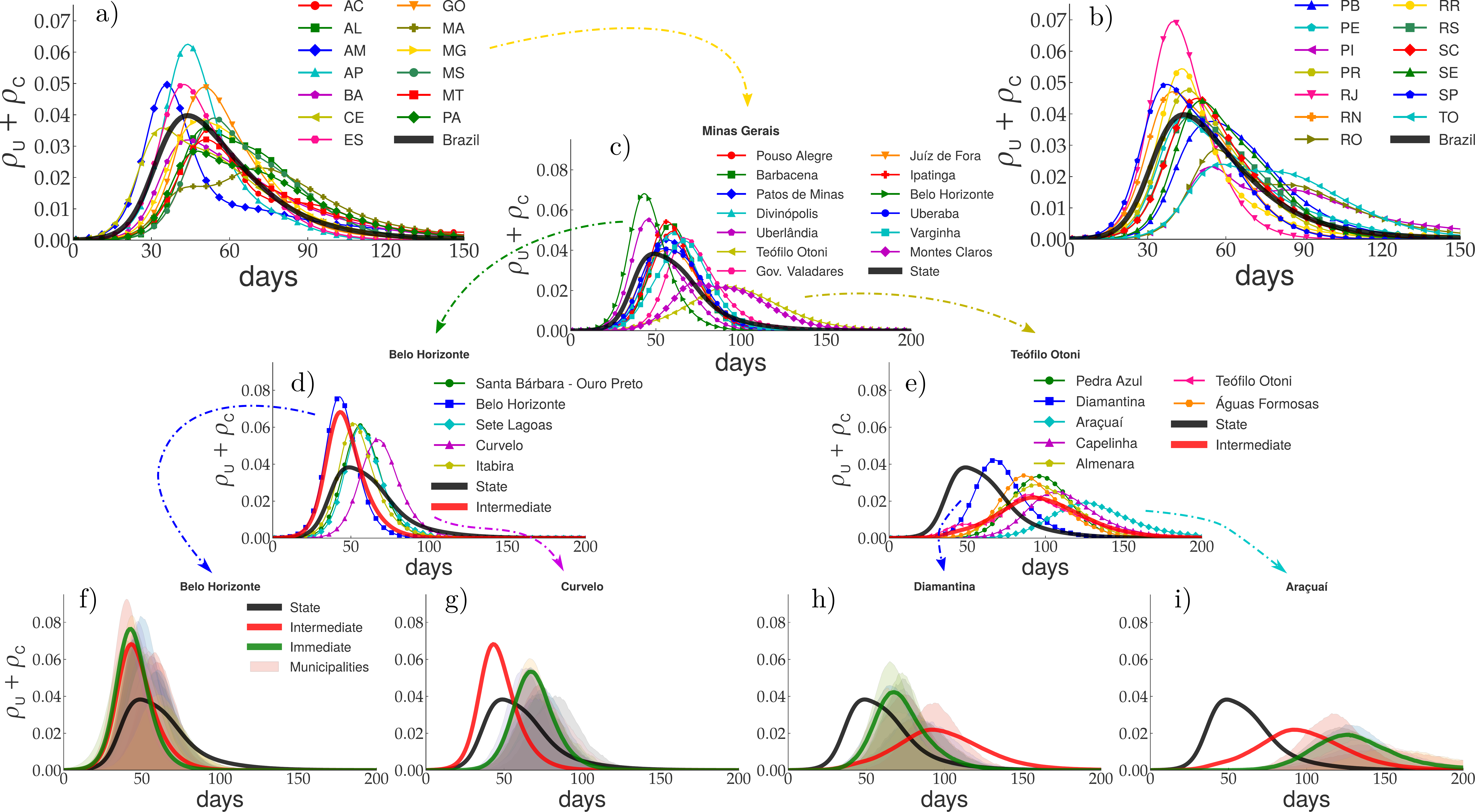}
	\caption{Multi-scale analysis of epidemic prevalence  at several scales of
		geographical organization, considering a weak mitigation with parameters
		$(M,K)=(0.4,0.3)$.   Epidemic curves  averaged with geographical resolution
		increasing from top to bottom are compared with lower resolution averages: a,b)
		States, c) intermediate regions, d,e) immediate regions, f-i) municipalities.
		The curves presenting the earliest and latest maxima are chosen as
		representative within each panel c) to e). Arrows indicate  curves selected for
		zooming. Day 0 corresponds to 31 March 2020. The state of MG was chosen for the
		higher resolution curves.}
	\label{fig:zoom_MG_R60K70}
\end{figure*}

From now on, generic  predictions under the aforementioned  hypothetical
scenarios are discussed. \add{We stress that these predictions, whose time window
	spans after the resubmission time,  were simulated in April 2020~\cite{Costa2020}}. High
variability of the epidemic outbreaks (intensity, duration and delays) can be
observed through all geographical scales.  Diversity was found in other
metapopulation approaches for epidemic processes~\cite{Danon2009,Danon2020} but
the multi-scale dispersion in our simulations is remarkable. At the state
level, the time when the outbreak reaches its maximum epidemic
prevalence, the epidemic peak, varies considerably as shown in
Figs.~\ref{fig:zoom_MG_R60K70}a) and \ref{fig:zoom_MG_R60K70}b), where the
epidemic prevalences for all states are shown as function of time for a weak
mitigation.

Let $T$ and $\rho$ be the  time and prevalence of the epidemic peak\add{, considering both unconfirmed and confirmed individuals}.
According to the simulations with weak mitigation, the first state to reach the
peak would be CE ($T_\text{CE}\approx34$~d), followed closely by SP and AM
($T_\text{SP}\approx T_\text{AM} \approx36$ d). The respective prevalences would
be $\rho_\text{CE}\approx3.6\%$, $\rho_\text{SP}\approx 4.9\%$, and
$\rho_\text{AM}\approx 5.0\%$. The state with highest epidemic prevalence would
be RJ with $\rho_\text{RJ} \approx 7.0\%$   at  $T_\text{RJ}\approx 40$~d. The
epidemic peak would happen later ($T_\text{MA} =71$ d), with maximum  less intense
($\rho_\text{MA}\approx 2.3$\%) and the curve  broader for the MA state. The MA
case is particularly interesting because whereas the peak for the state would
happen last, for its capital city São Luis it would occurs at day 41 slightly
earlier than the average peak for  Brazil, which would be at day $44$ in this
scenario. These numbers should not be quantitatively compared with the real
reported cases. Firstly, the confirmed cases correspond to  a small fraction of
the total of cases that depends on testing rate, an unknown variable in our study, and curves
for $\rho_\text{C}$ are much broader and have the peak postponed to later times.
Secondly, the uniform and constant mitigation and testing scenarios do not
correspond to reality. 

\begin{figure}[ht]
	\centering
	\begin{minipage}{0.950\linewidth}
		\includegraphics[width=0.99\linewidth]{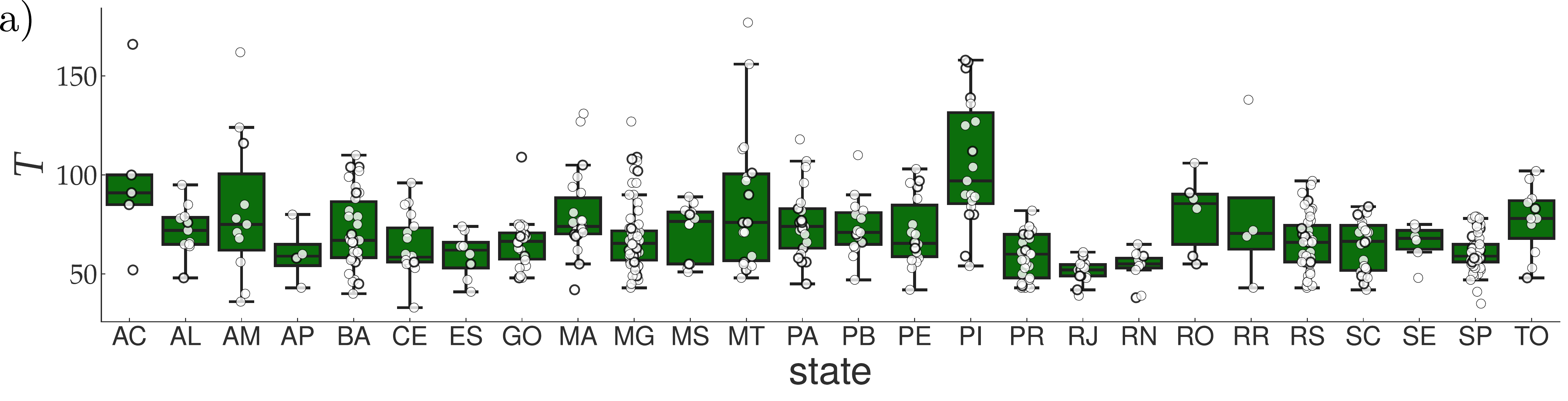}
		\includegraphics[width=0.99\linewidth]{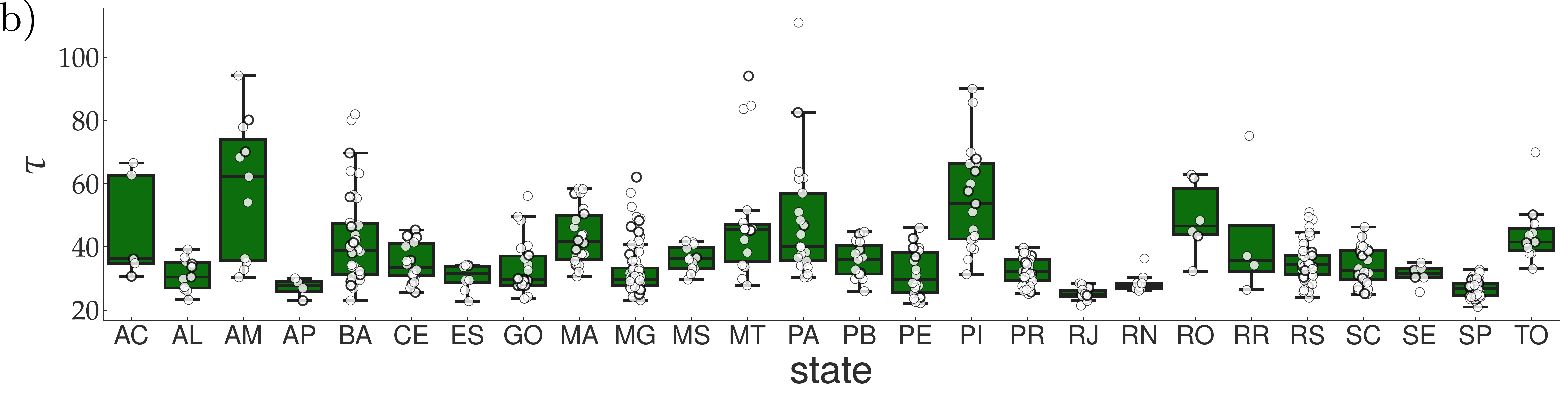}
		\includegraphics[width=0.99\linewidth]{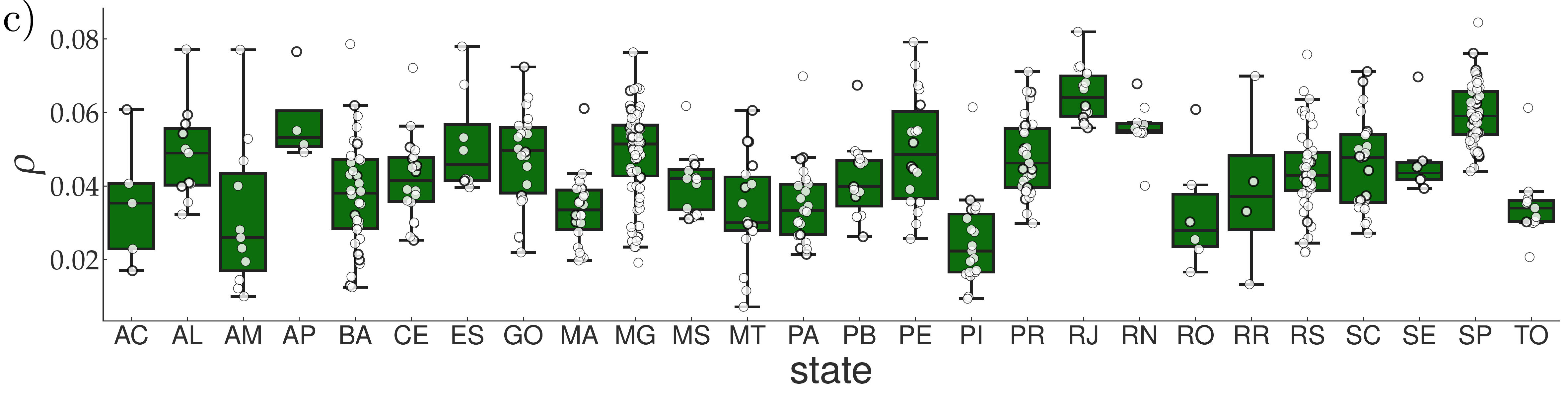}
	\end{minipage}
	\caption{Box plots for the peak a) time $T$, b) width $\tau$, and c) prevalence
		$\rho$ averaged over immediate regions for the 26 federative states using a
		weak mitigation with parameters $(M,K)=(0.4,0.3)$. Circles are data for
		different regions. As usual, boxes yield median, lower and upper quartiles
		while points outside whiskers are outliers.}
	\label{fig:boxplot}
\end{figure}
A spatially more  refined analysis is done comparing immediate and intermediate
geographical regions of a same federative state, which are determined by the
\textit{Instituto Brasileiro de Geografia e Estatística}
(IBGE)~\cite{instituto2017divisao}. An immediate region is a structure of nearby
urban areas with intense interchange for immediate needs such as purchasing of
goods, search for work, health care, and education. An intermediate region
agglomerates immediate regions preferentially with the inclusion of large
metropolitan areas~\cite{instituto2017divisao} laying thus between immediate
regions and federative states. In Brazil there are 133 intermediate and 510
immediate regions. Municipalities in different federative states do not belong to
the same immediate regions even if they have strong ties. The broad variability
of epidemic outbreaks observed  at the federative state level is repeated at
higher resolutions of geographical aggregation.
Figures~\ref{fig:zoom_MG_R60K70}c) to i) show the multi-scale structure of the
outbreaks within the state of MG. They compare the epidemic prevalences of
different intermediate regions, followed by the next scale with all immediate
regions  belonging to the pair of intermediate ones presenting the epidemic peak
first and last in Figs.~\ref{fig:zoom_MG_R60K70}d) and e). Finally, the highest
resolution with curves for municipalities within the respective immediate regions
with the earliest and latest maxima are shown in Figs.~\ref{fig:zoom_MG_R60K70}f)
to i).

A large dispersion through the 13 intermediate regions of the MG state is seen in
Fig.~\ref{fig:zoom_MG_R60K70}c). The times and prevalence values at the peak
differ by factors 2 and 6, respectively,  between the regions that attain the
maximum first and last. Curves for all intermediate regions of the 26 federative
states are  shown in Figs.~S1, S2,
and S3 of the {SM}~\cite{supp} for distinct mitigation
strategies, showing that the more intense the mitigation the more heterogeneous
are the curves across the country. Stepping down to immediate regions, a similar
behavior is found in Figs.~\ref{fig:zoom_MG_R60K70}d) and e). The intermediate
and immediate regions of Belo Horizonte, that include the capital city and are by
far the most populous regions of this state, lead the averages over  state and
intermediate regions, respectively, blurring the actual prevalence  in other
localities. Variability is still large  within immediate regions as shown in
Figs.~\ref{fig:zoom_MG_R60K70}f) to i) with the epidemic prevalence (shaded
plots) at different municipalities of a same immediate region. This pattern is
repeated for other federative states: only the dispersion amplitude varies. This
multi-scale nature for CE, RJ, RS,  and SP states are respectively provided in
Figs.~S4, S5, S6, and
S7  of the  SM~\cite{supp}.

Variability of  time, prevalence, and width $\tau$ (outbreak duration) of the
epidemic peak throughout immediate regions are shown in the box plots of
Fig.~\ref{fig:boxplot} for simulations with a  weak mitigation. Box plots for
none and moderate mitigations are qualitatively similar and presented in
Figs.~S8 and S9 of the SM~\cite{supp}.
One can see that federative states with smaller territorial areas, such as SE,
RN, and RJ, tend to have smaller dispersions of the peak time while those with
larger and more sparsely connected territories, as AM, MT, and MS, have larger
dispersions and outliers. We also have states with larger but better connected
territories such as SP, PR, and RS, for which the dispersion is significant but
not characterized by outliers. Irrespective of the more cohesive outbreak
patterns, the epidemic peaks in immediate regions of the SP state, for example,
could take  more than twice as long as the peak at the capital city. Conversely,
the outbreaks in the RJ state would be less desynchronized  than in its
neighbors, MG and SP, with the latest immediate region reaching the epidemic peak
only 50\% later than the capital city Rio de Janeiro, in this scenario.

\begin{figure*}[t]
	\centering
	\includegraphics[width=0.99\linewidth]{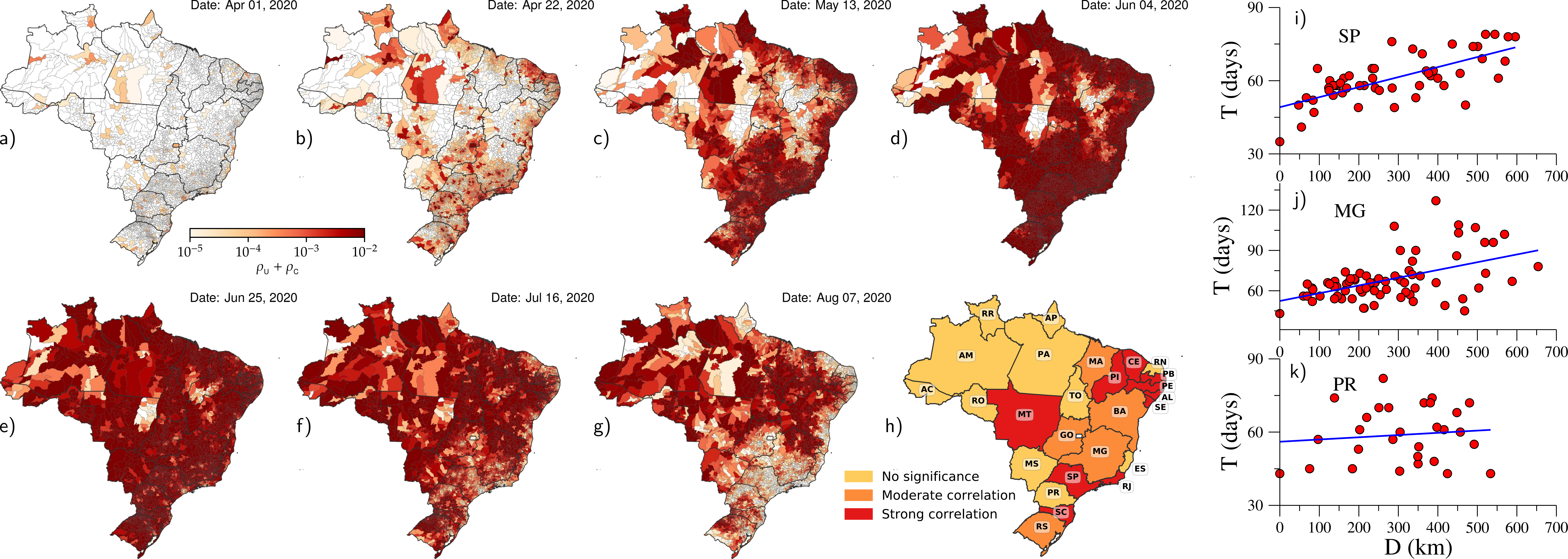}
	\caption{a-g) Color maps presenting the evolution of  the  prevalence of
		symptomatic cases (U and C) for Brazil in a simulation with a weak mitigation
		using parameters $(M,K)=(0.4,0.3)$. Dates of the simulation are shown in the
		upper right corner of each frame.  The darker colors represent higher
		prevalences in a logarithm scale. h) Map with the classification of typical
		correlation between peak time $T$ and distance of immediate regions to capital
		cities $D$ of each federative state. See main text for criteria. Scatter plots
		of $T$  versus $D$ for the  i) SP, j) MG and, k) PR states.}
	\label{fig:maps}
\end{figure*}

The space-time progression of the epidemic presents complex patterns as shown in
Figure~\ref{fig:maps}a) to g), where color maps with the prevalences of
symptomatic cases in each municipality of Brazil are shown with  intervals of
three weeks for simulations using weak mitigation; see 
Figs.~S10, and S11 of the 
SM~\cite{supp} for other scenarios. Full evolutions for all investigated 
mitigation parameters are provided in Movies S1,
S2, and  S3 of the  SM~\cite{supp}.  Epidemic
waves propagating on average from the coast at  east, where the most populated regions are,
to inland can be seen at a country level. Moreover, when the epidemic is
losing strength in the coast, the inland regions are still facing high levels of
epidemic prevalence. Maps and movies suggest that the epidemic starts in the more
populated places, usually the largest metropolitan regions, and spreads out
towards the countryside. This is quite clear in the SP state, where the focus  begins in
its capital city and moves westward into the countryside. However, other states
as PR and RS follow a different pattern with multiple foci evolving
simultaneously.

We analyzed the correlation between time $T$  of the epidemic peak in the
prevalence curves of immediate regions  and their distances $D$ from the capital
city of the respective state. Figures~\ref{fig:maps}i)-k) show  scatter plots of
$D$ vs $T$ for three typical correlation patterns found in the simulations. The
remaining states are given in Fig.~S12 of the SM~\cite{supp}.
The SP state presents a  strong correlation with Pearson coefficient $r=0.69,$
statistically significant with $p$-value less than $10^{-7}$. For PR, no
statistical correlation is observed. Finally, MG has a moderate correlation with
Pearson coefficient $r=0.47$, but statistically significant with $p$-value of
$10^{-5}$. We classified the states according to the correlation between $T$ and
$D$ in three groups: \textit{not significant} for which the $p$-value is larger
than $0.02$; \textit{strong correlation} that has statistical significance with
$p<0.02$ and $r>0.6$; and \textit{moderate correlation} with $p<0.02$ and
$r<0.6$. The results are summarized in Fig.~\ref{fig:maps}h)  where the map shows
states colored  according to their classification. The structure does not change
for other mitigation approaches. The complete set of Pearson coefficients and
$p$-values are given in Fig.~S12 of the SM~\cite{supp}.
States in the north (RR, RO, AP, AC, AM, PA, and TO) have few immediate regions
distributed in large territories and are less connected. Thus, lack of
correlations is not surprising. Moderate correlations in states as MG, BA, and
GO, with large but better connected territories are due to the existence of
multiple important regions that play the role of regional capitals. However, PR
is an extreme case with many and well connected immediate regions but  total lack
of correlations with the capital city, Curitiba. 
%Indeed,  PR has economically and
%culturally independent regions such as Londrina and Maringá with strong ties with
%SP, and the touristic region of Foz do Iguaçu with Paraguay and Argentina.

\section{Discussion}
\label{sec:discu}

The wide territorial, demographic, infrastructural diversities of large
countries as Brazil demand modeling of a pandemic  with geographical resolution
higher than the usual compartmental epidemic models, which can be achieved using
the metapopulation framework. We developed a stochastic metapopulation model and
applied it to the COVID-19 spreading in Brazilian municipalities. Performing
simulations with different hypothetical mitigation scenarios and considering the
integration of regions through \add{inter-municipality}
recurrent mobility, we have identified a high degree of heterogeneity and
desynchronization of the epidemic curves between the main metropolitan areas and
other inland regions. The diversity of outcomes is observed in several
geographical scales, ranging from states to immediate regions, that enclose
groups of nearby municipalities with strong economic and social ties. The
intenser the mitigation attitudes, the more remarkable are the differences.
Moderate or strong  correlations between the delay of the epidemic peak in the
interior regions and distance to the capital cities of the respective state were
observed for most states with well connected immediate regions. However, one
exception was the PR state where there are diverse epidemic foci  evolving
apparently independently of its capital city.

Our simulations suggest that uniform mitigation measures are not the optimal
strategy. In a municipality where the peak  without interventions would happen
quite later than the capital city of its state, it would delay even more if
strong mitigation measures are adopted synchronously with the capital.
Nevertheless, such a place probably would have to extend the mitigation for much
longer periods since once the epicenters of epidemics start to relax their
restrictions, the viruses will circulate fast, reaching these vulnerable
municipalities with most of  population still susceptible. \add{This is the actual
	current situation in Brazil at moment of resubmission.} The social and economic  impacts
could be higher. On the other hand, an eventual collapse of the local heath 
system, a central question of the COVID-19 pandemic combating~\cite{Arenas}, in
the countryside regions could also occur after the larger metropolitan areas
were under control, raising the possibility of unburdening the local healthy
care system. In fact, the desynchronization of the epidemic curves and
consequently of the health-system collapse could be an asset in designing
optimal resource allocations. One could prevent rapid expansion of the epidemic
through smaller municipalities flattening the curves as much as possible in the
main metropolitan regions. Beyond the positive consequences for their own health
systems, this attitude would have the positive side effect of  delaying
outbreaks in the countryside. Our study shows that countryside regions are not
safe in the medium-term but could have precious additional time to prepare
themselves.

{\color{black}
\section{Afterwords}
\label{sec:after}

There is an interval of five months between conceiving the first version of the
present work~\cite{Costa2020} and this final submission. As emphasized
throughout the paper, our aim  was to simulate hypothetical scenarios  rather to
do an accurate forecasting of real epidemic curves. Therefore,  qualitative
behaviors predicted in the simulations can now be confronted with the real
series observed \textit{a posteriori}.  Several qualitative aspects predicted in
simulations were indeed observed in real data. Maybe the most remarkable one is
the confirmation of  the multi-scale nature of the epidemic curves presented in
Fig.~\ref{fig:zoom_MG_R60K70} of the main text and Figs.~S4 to S7 of
SM~\cite{supp}. Figure~\ref{fig:real} presents the epidemic incidences (new
cases per capita over a week) for intermediate regions of three states
presenting  typical behaviors observed in simulations. The scenario of
propagation starting from east coast towards inland, with a considerable delay
for the epidemic peak in the latter, shown in Fig.~\ref{fig:maps} was also
observed.  Other interesting phenomena were also verified in real data as, for
example, the unusual curve  for the MA state and the multifocal spreading in
some states as PR, MG and GO.

Lets consider in more details  the SP state shown in Fig~\ref{fig:real}a), the
one with highest number of COVID-19 cases in Brazil. One can see a moderate
dispersion for  peak occurrence time  while in the capital city São Paulo it
happened considerably earlier. This picture is  the same drawn in Fig.~S7a) of
the SM~\cite{supp}. For the state of RJ, shown in Fig.~\ref{fig:real}b), the
real data, with an exception for the intermediate region of Petropólis, are remarkably similar to
simulations (see Fig.~S5a) of the SM~\cite{supp}), where intermediate regions
peaks approximately synchronously with the intermediate region of Rio de Janeiro. We are not able to
provide a specific reason for the different behavior of Petropólis. The
incidence curves for the CE state, one of the first and more intensely  impacted
places by COVID-19 in Brazil and shown in Fig. \ref{fig:real}c), are remarkably
heterogeneous with  Fortaleza reaching the maximum much earlier and Cretéus much
later than the other intermediate regions. The incidence curve for the whole CE
state is very broad  meaning that different regions rule the state'
curve at different times, a scenario in almost perfect consonance with the
simulation predictions shown in Fig. S4 of SM~\cite{supp}. We can also see 
significant dispersion for  immediate regions' curves   within a same
intermediate region, as illustrated for
Sobral in Fig.~\ref{fig:real}d).  Intermediate region of Cretéus is composed only
by two intermediate regions and does not illustrate the point as well as Sobral
does. The maximum incidence  also varied considerably
as, once more, predicted in simulations.

Quantitatively, a hypothetical scenario fitting better more regions lies 
between weak and moderate mitigations. However, as one
could  guess, different hypothetical scenarios works better at distinct
moments and places confirming that quantitative forecasting and even nowcasting 
is a very dedicated issue.

\begin{figure*}[th]
\centering
\includegraphics[width=0.7\linewidth]{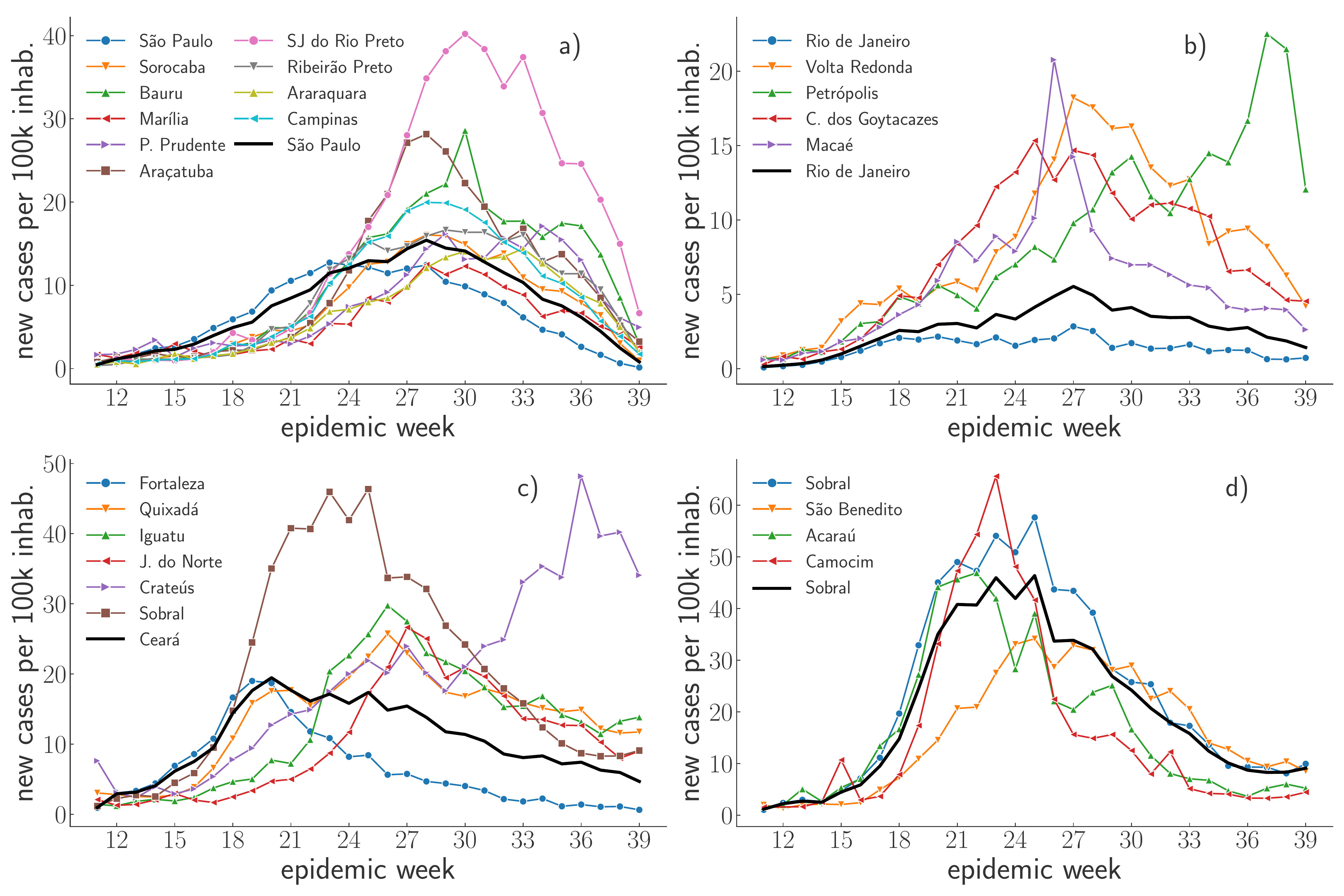}
\caption{\color{black}Real time series of epidemic incidence per week  for
	intermediate regions of the states of  a) SP, b) RJ, and c) CE and for the d)
	immediate regions belonging to the intermediate region of Sobral in the state of
	CE. Thick lines are the incidence for lower resolution scale in the
	corresponding a)-c) state or d) intermediate region. Data reported by onset of
	symptoms were extracted from \textit{e-SUS NOTIFICA} official dataset held by
	Brazilian Ministry of Health~\cite{eSUS2020}. Epidemic week 12 correspond to
	March 15-21, 2020.}
\label{fig:real}
\end{figure*}

}

\appendix

\section{Computer implementation}

\label{app:model}

The computer implementation of the epidemic model follows the Gillespie methods based on Ref.~\cite{Cota}.

%\subsection
{\it Definitions.}
We define $n^\text{X}_{ji}$ as the number of individuals in state X (= S,E,A,U,C,
or R) within the patch $i$ that comes from patch $j$ such that 
$\tilde{n}^\text{X}_{i} = \sum_j n^\text{X}_{ji}$  is the number of individuals
in  patch $i$ that are in state X. Thus, the number of residents of $i$  in
state X at patch $i$ is  $n^\text{X}_{ii}$  and of nonresidents is given by
$\tilde{n}^\text{X}_{i}- n^\text{X}_{ii}$. Also,  $\tilde{n}_i=\sum_{X} 
\tilde{n}^\text{X}_{i}$ is the total number  of individuals and $\tilde{n}_{ii}=\sum_{X} 
\tilde{n}^\text{X}_{ii}$ the number of residents in the patch $i$. Finally,
the total number of individuals in the compartment X is
$N_\text{X} = \sum_{j=1}^{\Omega}\tilde{n}^\text{X}_{i}$.

We have that $W_{ii} = N_i - \sum_{j\neq i} W_{ij}$ is the
population of patch $i$ that does not leave it. 
The population  fraction moving recurrently from $i$ to $j$ is
\begin{equation}
R_{ij}=\frac{W_{ij}}{\sum_{l} W_{il}}.
\end{equation}
Thus, $R_i=\sum_{j\ne i} R_{ij}$ and $R_{ii}=1-R_i$ are, respectively, the
population fractions  that move or not from patch $i$.

%\subsection
{\it General structure of the simulation.}
We divide the simulation in two big groups: mobility and epidemic events. The
total rate of \add{inter-patch} mobility is
\begin{equation}
D_\text{tot}=T_\text{r}^{-1}\sum_{i} \tilde{n}_{ii} R_{i} +T_\text{r}^{-1}\sum_{i}(\tilde{n}_{i}-\tilde{n}_{ii}),
\end{equation} 
in which the first term represents the individuals leaving from and the second
returning to their patches of residence.

The epidemic transitions $\mathcal{T}$ and their respective total rates $\Gamma_\mathcal{T}$  are defined in Table~\ref{tab:Trates}.
\begin{table}[ht]
	\caption{Total transition rates for different types of epidemic events.}
	\label{tab:Trates}
	\begin{center}
		\begin{tabular}{ccccccc}
			\hline\hline
			$\mathcal{T}$ &  ~~~&  $\Gamma_\mathcal{T}$ & ~~~ & $\mathcal{T}$ &  ~~~&  $\Gamma_\mathcal{T}$\\\hline
			E$\rightarrow$A & & $\mu_\text{A} N^\text{E}$ & &
			A$\rightarrow$U & & $\beta_\text{U} N^\text{A}$\\
			A$\rightarrow$R & & $\beta_\text{R}  N^\text{A}$& &
			U$\rightarrow$R & & $\alpha_\text{R} N^\text{U}$\\
			U$\rightarrow$C & & $\alpha_\text{C} N^\text{U}$& &
			C$\rightarrow$R & & $\alpha_\text{R} N^\text{C}$\\
			S$\rightarrow$E & & %$\sum_i \Gamma^{\text{S}\rightarrow\text{E}}_{i}$
			$\sum_{i=1}^{\Omega} \Theta_i$  & & & & \\
			\hline\hline
		\end{tabular}
	\end{center}
\end{table}

{The infection rate in a patch $i$ is given by 
	\begin{equation}
	\Theta_i = \dfrac{k_i\tilde{n}^\text{S}_{i}}{\tilde{n}_i} \left(\lambda_\text{A}\tilde{n}^\text{A}_{i}+
	\lambda_\text{U}\tilde{n}^\text{U}_{i} + b\lambda_\text{C}\tilde{n}^\text{C}_{i} \right),
	\end{equation}
	%Here $\Gamma^{\text{S}\rightarrow\text{E}}_i=  
	%\av{k}\tilde{n}^\text{S}_{i}\left(\lambda_\text{A}\tilde{n}^\text{A}_{i}+
	%\lambda_\text{I}\tilde{n}^\text{I}_{i}\right)$ 
	while the total rate of epidemic events is 
	$
	%\begin{equation}
	E_\text{tot} = \sum_{\mathcal{T}} \Gamma_\mathcal{T}.
	%(\mu_\text{A+}\mu_\text{I})N_\text{E}+(\beta_\text{R}+\beta_\text{I})N_\text{A}+
	%\alpha N_\text{I}+\av{k}\sum_{i=1}^{\Omega}\tilde{n}^\text{S}_{i}\left(\lambda_\text{A}\tilde{n}^\text{A}_{i}+
	%\lambda_\text{I}\tilde{n}^\text{I}_{i}\right),
	$
	%\end{equation}
	In each time step, we select  one mobility or epidemic event  with probabilities
	$P_\text{mob}=D_\text{tot}/(D_\text{tot}+E_\text{tot})$ and $P_\text{epi}=1-P_\text{mob}$,
	respectively, and the time is increased by $\delta t=-\av{\delta t} \ln\eta$ where $\av{\delta t} = 1/(D_\text{tot}+E_\text{tot})$ and $\eta$ is a pseudo random number uniformly distributed in the interval (0,1].
	
%	\subsection
	{\it \add{Inter-patch} mobility events.}
	If a mobility event was chosen we proceed as follows.
	\begin{itemize}[leftmargin=*]
		\item [i)] {Choose} the patch $i$ from where an individual will leave proportionally to the
		total mobility rate of the patch, which occurs with probability
		\begin{equation}
		Q_i= \frac{\tilde{n}_{ii}R_i + (\tilde{n}_{i}-\tilde{n}_{ii})}{\sum_j \left[\tilde{n}_{jj}R_j + (\tilde{n}_{j}-\tilde{n}_{jj})\right]}
		\end{equation}

		\item [ii)] Choose the state X of the individual  of patch $i$ that
		will move with probability
		\begin{equation}
		P_{i}^{\text{X}} = \frac{\tilde{n}_{i}^\text{X}}{\sum_\text{Y}\tilde{n}_{i}^\text{Y}}. 
		\end{equation}
		\item[iii)] Select if the individual to move is resident or nonresident with probabilities
		$n^\text{X}_{ii}/\tilde{n}^\text{X}_{i}$ and $1-n^\text{X}_{ii}/\tilde{n}^\text{X}_{i}$, respectively.
		
		\item[iv)] If the individual is resident and not symptomatic (S,E,A or R), the
		destination patch $j$ is chosen with probability $R_{ij}/\tilde{R}_{i}$. If he/she is symptomatic, stays in the resident patch ($j=i$). The movement
		is implemented as $n^\text{X}_{ij}\rightarrow n^\text{X}_{ij}+1$ and 
		$n^\text{X}_{ii}\rightarrow n^\text{X}_{ii}-1$.

		\item[v)] If the individual to move is a nonresident, one patch $j$ is chosen
		with probability $n_{ji}^\text{X}/(\tilde{n}_{i}^\text{X} -
		\tilde{n}_{ii}^\text{X})$ and the movement is implemented as
		$n_{ji}^\text{X}\rightarrow n_{ji}^\text{X}-1$ and $n_{jj}^\text{X}\rightarrow
		n_{jj}^\text{X}+1$.
	\end{itemize}
	
%	\subsection
	{\it Epidemic events.}
	If an epidemic event was chosen we proceed as follows.
	
	\begin{itemize}[leftmargin=*]
		\item[i)] One of the transitions $\mathcal{T}$ is chosen with probability
		\begin{equation}
		P_\mathcal{T}=\frac{\Gamma_\mathcal{T}}{\sum_{\mathcal{T}}\Gamma_\mathcal{T}}.
		\end{equation}
		\item[ii)] If $\mathcal{T}=$~S$\rightarrow$E, then we choose a patch $i$ with
		probability $\Theta_i/\sum_{j=1}^{\Omega}\Theta_j$. Then, one  patch $j$ is chosen with
		probability $n^\text{S}_{ji}/\tilde{n}^\text{S}_i$ and the populations are updated as 
		$n^\text{S}_{ji}\rightarrow n^\text{S}_{ji}-1$ and $n^\text{E}_{ji}\rightarrow
		n^\text{E}_{ji}+1$. Notice that both resident ($i=j$) or nonresident ($i\ne j$)
		are considered in this rule.
		
		\item[iii)] If $\mathcal{T}$ is a spontaneous transition of the form 
		X$\rightarrow$Y then a patch $i$ is chosen with probability
		$\tilde{n}^\text{X}_i/N^\text{X}$ and the patch $j$ of residence is chosen with
		probability $n^\text{X}_{ji}/\tilde{n}^\text{X}_i$. Then
		$n^\text{X}_{ji}\rightarrow n^\text{X}_{ji}-1$ and $n^\text{Y}_{ji}\rightarrow
		n^\text{Y}_{ji}+1$.

	\end{itemize}
	
	This algorithm spends too much time with mobility of susceptible and removed
	individuals since they can exist in very large number for long times. We
	implemented an optimization performing the mobility of susceptible and removed
	in discrete times of $0.5~$d. So, instead of moving one susceptible individual
	at a time step, we move $n_{ii}^\text{S}R_{ij}$ from $i$ to $j$ at once and
	those who were not infected while visiting patch $j$, return to $i$  after
	$0.5~$d. The same is done for removed individuals. The simulations yield the
	same results as the exact procedure within the statistical uncertainties.

\section{Parameters for a data-driven approach}
\label{app:par}

%\subsection{Mobility}
{\it Mobility.} Mobility parameters were extracted from public sources. Local
commuting $C_{ij}^{(1)}$  was estimated from the national census data of 2010
provided by IBGE~\cite{instituto2012censo} as the number of people that daily
commute among municipalities to work or study. We excluded from the dataset
those travels corresponding to more than 250~km, which are reckoned by  airline
travel data. Commuting data lacking destinations were included, assuming
proportionality to the complete-information portion of the dataset;
see~\cite{Lana2017} for details. For the sake of generality, the recurrent flux
is a function $W_{ij}^{(1)}=F_{ij}(C_{ij}^{(1)})$. We  use the simplest case
$F_{ij}(x) =x$. Air transportation data were obtained from the~\textit{Agência
	Nacional de Aviação Civil} (ANAC)~\cite{ANAC} with the statistics of direct
flights of 2014 considering the main nearby metropolitan region as the origin
and destination of passengers. Using boarding and landing statistics of
passengers, one-connection flights were inferred and the number of passengers
$C_{ij}^{(2)}$ flying daily from $i$ to $j$ was estimated~\cite{Coelho2020}. In
order to construct a recurrent mobility with flights, we assume
$W_{ij}^{(2)}=(C_{ij}^{(2)}+C_{ji}^{(2)})/2$, which represents an effective
flux. \add{We do not have estimates for duration of long-range travels. However,
	this does not affect the current model since } the homogeneous mixing hypothesis
within the patches means that only the  number of individuals that stay part of
the time in two different patches actually matters  and not who traveled. The
total recurrent mobility is $W_{ij}=W_{ij}^{(1)}+W_{ij}^{(2)}$; we assume that
it occurs daily, i.e, $T_\text{r}=1$~d \add{since both commuting and air travel
	data were obtained with this time resolution}.

%\subsection
{\it Epidemiological parameters.}
We used epidemiological parameters based on Ref.~\cite{Arenas} which were mined
from  analyses of early COVID-19 epidemic spreading on Hubei province
China~\cite{Li2020a,Danon2020,Read2020}. The total time  in the exposed and
asymptomatic compartments was taken  as $\mu_\text{A}^{-1}+\beta_\text{U}^{-1}
=5.2~\text{d}$~\cite{Li2020a}. We used $\mu_\text{A}^{-1} =
\beta_\text{U}^{-1}=2.6$~d. The  time for a symptomatic individual be removed
was estimated as $\alpha_R^{-1}=3.2~\text{d}$~\cite{Read2020,Danon2020}. For the
\add{truly asymptomatic individuals, we used the same recovering time of 
the presymptomatic ones}, such that $\beta_\text{R}^{-1} =
\beta_\text{U}^{-1}+\alpha^{-1}_\text{R}$. The infection rates $\lambda_\text{C}
= \lambda_\text{U} = \lambda_\text{A} =
0.06$~d$^{-1}$~\cite{Chinazzi2020,pitzer2007estimating,Arenas} and average
number of contacts $\av{k}=13$ were estimated to reproduce the scaling of
Brazilian very early exponential growth, $\sim\exp{(0.3t)}$, in reported cases of
COVID-19; see Fig.~S13 in the SM~\cite{supp}. These parameters
provide an effective infection rate $\lambda\av{k}=0.78$ that represents an
optimistic estimate nearer to the lower bounds estimated in other studies of
COVID-19~\cite{Read2020,Zhan2020,Arenas,Danon2020}. The contact of the
individuals who were confirmed for COVID-19 was depleted to a fraction $b=0.3$
of their regular contacts. The confirmation rate $1/\alpha_\text{C}^{(0)}=10$~d
was calibrated together with the initial conditions  to fit the amplitude of
confirmed case curves after 31 March 2020 and also the total number of
municipalities with confirmed cases in a weak mitigation scenario. \add{This is approximately 
	twice the incubation time implying that symptomatic individuals may not be tested
	even if tests are available. } Up to 23
April 2020, Brazil had performed less than $1400$ tests per million
inhabitants~\cite{covidBR}. Assuming a time window of 35~d for testing, we
have approximately $\zeta=1/40~000$ d$^{-1}$ per inhabitant. Due to the testing
targeted only to severe cases of respiratory syndromes adopted in Brazil to the 
date of investigation, the transition from asymptomatic to  confirmed was
neglected setting $\beta_\text{C}=0$;  note  that the model can  easily include
such a transition if the testing scenario changes. 
%Since our work is focused at
%hypothetical scenarios rather than forecasting of the real situation, these
%limitations on testing data are not major concerns of the present work but can
%certainly be in forthcoming forecasting studies based on the present model.

%\subsection
{\it Demographic parameters.}
Each patch represents a municipality with resident population given by  IBGE
using estimates of 2019~\cite{ibge2019estimativas}, while the respective
fractions of rural and urban populations  were obtained from the IBGE census of
2010~\cite{instituto2012censo}. The population density was determined using
urban areas,  obtained from  high resolution satellite images, provided by
\textit{Empresa Brasileira de Pesquisa Agropecuária}
(EMBRAPA)~\cite{farias2017identificaccao}. Both fraction and density of urban
population, and, consequently, the number of contacts have broad distributions
as can be seen in Fig.~S13 of the SM~\cite{supp}. These are key
features to enhance the accuracy of the predictions beyond large metropolitan
centers.  Parameter $a$ that controls the function $f(x)$ was taken as
$a=1/\av{\xi}$ where $\av{\xi}=2791$~inhabitants/km$^2$ is the average urban
population density  of Brazil. The rural-urban contact matrix was chosen as
$\mathcal{C}_\text{uu}=\mathcal{C}_\text{rr}=1$ and $\mathcal{C}_\text{ru}=0.5$.
This last choice only assumes that urban-rural contacts are significantly
smaller than urban-urban or rural-rural interactions.

\begin{acknowledgments}
We thank Marcelo F. C. Gomes for providing  the filtered mobility data
used in the present work; Ronald Dickman for useful discussion and manuscript
reading; Jesus Góméz-Gardeñes and Alex Arenas for discussions that motivated
the beginning of this work. WC thanks David Soriano-Paños for extensive
discussions about metapopulation modeling and Mario Balan for discussions about
IBGE data. WC thanks the kind hospitality of University of Zaragoza and GOTHAM
Lab during the realization of the work. 
%	SCF is in profound debit with Zélia
%	Almeida who kept two children calm and safe under the same roof where this work
%	was performed during the quarantine. 
\revis{This work was partially supported by the Brazilian agencies \textit{Coordenação
	de Aperfeiçoamento de Pessoal de Nível Superior} - CAPES (Grant no. 88887.507046/2020-00),
\textit{Conselho Nacional de Desenvolvimento Científico e Tecnológico}- CNPq
(Grants no. 430768/2018-4 and 311183/2019-0), and \textit{Fundação de Amparo à
	Pesquisa do Estado de Minas Gerais} - FAPEMIG (Grant no. APQ-02393-18).} This
study was financed in part by the \textit{Coordenação de Aperfeiçoamento de
	Pessoal de Nível Superior} (CAPES) - Brasil  - Finance Code 001.
\end{acknowledgments}

%\bigskip
%
%\noindent \textbf{Author contributions:}
%GSC performed the stochastic simulations. WC and GSC prepared the figures and
%movies. SCF wrote the first version of the manuscript. All authors contributed
%equally to result analysis, manuscript revision, and  research conception.
%
%
%\bigskip
%
%\noindent \textbf{Competing interests:} All authors declare no competing interest.
%
%\bigskip
%
%
%\noindent \textbf{Data availability:} All data used in our simulations are available
%in public repositories~\cite{covidBR,instituto2012censo,ANAC,farias2017identificaccao}. 
%
%\bigskip
%
%
%\noindent \textbf{Corresponding author:} Silvio C. Ferreira (silviojr@ufv.br).

%\bibliography{covid19_rev}

\end{document}